\documentstyle[pra,aps,multicol]{revtex}

\renewcommand{\narrowtext}{\begin{multicols}{2} \global\columnwidth20.5pc}
\renewcommand{\widetext}{\end{multicols} \global\columnwidth42.5pc}
\begin{document}
\draft
\title{Realizing probabilistic identification and cloning of quantum states via
universal quantum logic gates}
\author{Chuan-Wei Zhang, Zi-Yang Wang, Chuan-Feng Li\thanks{%
Electronic address: cfli@ustc.edu.cn}, and Guang-Can Guo\thanks{%
Electronic address: gcguo@ustc.edu.cn}}
\address{Laboratory of Quantum Communication and Quantum\\
Computation and Department of Physics,\\
University of Science and Technology of China,\\
Hefei 230026, People's Republic of China}
\maketitle

\begin{abstract}
\baselineskip12ptProbabilistic quantum cloning and identifying machines can
be constructed via unitary-reduction processes [Duan and Guo, Phys. Rev.
Lett. {\bf 80}, 4999 (1998)]. Given the cloning (identifying) probabilities,
we derive an explicit representation of the unitary evolution and
corresponding Hamiltonian to realize probabilistic cloning (identification).
The logic networks are obtained by decomposing the unitary representation
into universal quantum logic operations. The robustness of the networks is
also discussed. Our method is suitable for a $k$-partite system, such as
quantum computer, and may be generalized to general state-dependent cloning
and identification.

PACS number(s): 03.67.-a, 03.65.Bz, 89.70.+c
\end{abstract}

\vskip 1.0cm

\narrowtext
\baselineskip11pt

\section{Introduction}

Quantum no-cloning theorem \cite{Woot82}, which asserts that unknown pure
states cannot be reproduced exactly by any physical means, is one of the
most astonishing features of quantum mechanics. Wootters and Zurek \cite
{Woot82} have shown that the cloning machine violates the quantum
superposition principle. Yuen and D'Ariano \cite{Yuen86,Yuen96} showed that
a violation of unitarity makes the cloning of two nonorthogonal states
impossible. Barnum {\it et al.} \cite{Bar96} have extended such results to
the case of mixed states and shown that two noncommuting mixed states cannot
be broadcast. Furthermore, Koashi and Imoto \cite{Koa98} generalized the
standard no-cloning theorem to the entangled states. The similar problem
exists in the situation of identifying an arbitrary unknown state \cite
{Die82}. Since perfect quantum cloning and identification are impossible,
the inaccurate cloning and identification of quantum states have attracted
much attention with the development of quantum information theory .

The inaccurate cloning and identification may be divided into two main
categories: deterministic and probabilistic. The deterministic quantum
cloning machine generates approximate copies and further we get two
subcategories: universal and state-dependent. Universal quantum cloning
machines, first addressed by Bu\v zek and Hillery \cite{Buz96}, act on any
unknown quantum state and produce approximate copies equally well. The Bu\v z%
ek-Hillery cloning machine has been optimized and generalized in Refs. \cite
{Gis97,Bru98A,Bru98L,Wer98,Key99}. Massar and Popescu \cite{Mas95} and Derka 
{\it et al. }\cite{Der98}{\it \ }have also considered the problem of
universal states estimation, given $M$ independent realizations. The
deterministic state-dependent cloning machine, proposed originally by
Hillery and Bu\v zek \cite{Hil98}, is designed to generate approximate
clones of states belonging to a finite set. Optimal results for two-state
cloning have been obtained by Bru$\beta $ {\it et al. }\cite{Bru98A} and
Chelfes and Barnett \cite{Che99}. Deterministic exact cloning violates the
no-cloning theorem, thus faithful cloning must be probabilistic. The
probabilistic cloning machine was first considered by Duan and Guo \cite
{Dua98A,Dua98L} using a general unitary-reduction operation with a
postselection of the measurement results. They showed that a set of
nonorthogonal but linear-independent pure states can be faithfully cloned
with optimal success probability. Recently, Chelfes and Barnett \cite{Che99}
presented the idea of hybrid cloning, which interpolates between
deterministic and probabilistic cloning of a two-state system. In addition,
we \cite{Zha99} have provided general identifying strategies for
state-dependent system.

Clearly, it is important to obtain a physical means to carry out this
cloning and identification. Quantum networks for universal cloning have been
proposed by Bu\v zek {\it et al. }\cite{Buz97}. Chelfes and Barnett \cite
{Che99} have constructed the cloning machine in a two-state system.

In this paper we provide a method to realize probabilistic identification
and cloning for an $n$-state system. The method is also applicable to
general cloning and identification of state-dependent systems. As any
unitary evolution can be accomplished via universal quantum logic gates \cite
{Deu85,Bar95}, the key to realizing probabilistic identification and cloning
is to obtain the unitary representation or the Hamiltonian of the evolution
in the machines. We derive the explicit unitary representation and the
Hamiltonian which are determined by the probabilities of cloning or
identification. Furthermore, we obtain the logic networks of probabilistic
clone and identification by decomposing the unitary representation into
universal quantum logic operations. The robustness of the networks is also
discussed.

The plan of the paper is the following. In Sec. II we derive the unitary
representation matrix and Hamiltonian for quantum identification provided
with one copy and generalize this method to $M\rightarrow N$ quantum cloning
and identification with $M$ initial copies. For the special case of a
quantum computer, we should be concerned with the system which includes $k$
partites, each of them being an arbitrary two-state quantum system (qubit).
The identification and cloning in such $k$-partite quantum systems have more
prospective applications, which include normal qubits and multipartite
entangled states. In Sec. III, we provide the networks of probabilistic
cloning and identification of $k$-partite system and discuss their stability
properties.

\section{Unitary evolutions and Hamiltonians for identification and clone}

Any operation in quantum mechanics can be represented by a unitary evolution
together with a measurement. Considering the states secretly chosen from the
set $S=\left\{ \left| \psi _i\right\rangle ,i=1,2,...,n\right\} $ which span
an $n$-dimensional Hilbert space, Duan and Guo \cite{Dua98L} have shown that
these states can be probabilistically cloned by a general unitary-reduction
operation if and only if $\left| \psi _1\right\rangle $, $\left| \psi
_2\right\rangle $,..., $\left| \psi _n\right\rangle $ are
linear-independent. By introducing a probe $P$ in an $n_P$-dimensional
Hilbert space, where $n_P\geq n+1$, the unitary evolution $\hat U$ in the $%
M\rightarrow N$ probabilistic cloning machine can be written as follows: 
\begin{eqnarray}
&&\hat U\left| \psi _i\right\rangle ^{\otimes M}\left| \varphi
_1\right\rangle ^{\otimes (N-M)}\left| P_0\right\rangle  \eqnum{2.1} \\
&=&\sqrt{\gamma _i}\left| \psi _i\right\rangle ^{\otimes N}\left|
P_i\right\rangle +\sum_jc_{ij}\left| \alpha _j\right\rangle \left| \varphi
_1\right\rangle ^{\otimes (N-M)}\left| P_0\right\rangle \text{,}  \nonumber
\end{eqnarray}
where $\left| P_0\right\rangle $ and $\left| P_i\right\rangle $ are
normalized states of the probe system (not generally orthogonal, but each of 
$\left| P_i\right\rangle $ is orthogonal to $\left| P_0\right\rangle $), and 
$\left| \psi _i\right\rangle ^{\otimes M}=\left| \psi _i\right\rangle
_1\left| \psi _i\right\rangle _2\cdots \left| \psi _i\right\rangle _M$ ($%
\left| \psi _i\right\rangle _k$ is the $k$th copy of state $\left| \psi
_i\right\rangle $). The $n$-dimensional Hilbert spaces spanned by state sets 
$\left\{ \left| \psi _i\right\rangle \right\} $, $\left\{ \left| \psi
_i\right\rangle ^{\otimes M}\right\} $, or $\left\{ \left| \psi
_i\right\rangle ^{\otimes N}\left| P_i\right\rangle \right\} $ are denoted
by ${\cal H}$, ${\cal H}^M$ and ${\cal H}^N$ respectively, and $\left\{
\left| \varphi _i\right\rangle \right\} $, $\left\{ \left| \alpha
_i\right\rangle \right\} $, and $\left\{ \left| \beta _i\right\rangle
\right\} $ are the orthogonal bases of each space. The probe $P$ is measured
after the evolution. With probability $\gamma _i$, the cloning attempt
succeeds and the output state is $\left| \psi _i\right\rangle ^{\otimes N}$
if and only if the measurement result of the probe is $\left|
P_i\right\rangle $. The $n\times n$ inter-inner products of Eq. (2.1) yield
the matrix equation 
\begin{equation}
X^{(M)}=\sqrt{\Gamma }X_P^{(N)}\sqrt{\Gamma }+CC^{\dagger },  \eqnum{2.2}
\end{equation}
where the $n\times n$ matrices are $C=\left[ c_{ij}\right] $, $X^{\left(
M\right) }=\left[ \left\langle \psi _i|\psi _j\right\rangle ^M\right] $, and 
$X_P^{(N)}=\left[ \left\langle \psi _i|\psi _j\right\rangle ^N\left\langle
P_i|P_j\right\rangle \right] $. The diagonal efficiency matrix $\Gamma $ is
defined as $\Gamma =%
\mathop{\rm diag}
(\gamma _1,\gamma _2,...,\gamma _n)$. Since $CC^{\dagger }\geq 0$ ($%
CC^{\dagger }$ is positive semidefinite), Eq. (2.2) yields 
\begin{equation}
X^{(M)}-\sqrt{\Gamma }X_P^{(N)}\sqrt{\Gamma }\geq 0\text{. }  \eqnum{2.3}
\end{equation}
This inequality determines the optimal cloning efficiencies. For example,
when $n=2$, we get \cite{Che99} 
\begin{eqnarray}
\frac{\gamma _1+\gamma _2}2 &\leq &\max_{\left\{ \left| P_i\right\rangle
\right\} }\frac{1-\left| \left\langle \psi _1|\psi _2\right\rangle \right| ^M%
}{1-\left| \left\langle \psi _1|\psi _2\right\rangle \right| ^N\left|
\left\langle P_1|P_2\right\rangle \right| }  \eqnum{2.4} \\
&=&\frac{1-\left| \left\langle \psi _1|\psi _2\right\rangle \right| ^M}{%
1-\left| \left\langle \psi _1|\psi _2\right\rangle \right| ^N}\text{.} 
\nonumber
\end{eqnarray}

In the limit as $N\rightarrow \infty $, the $M\rightarrow N$ probabilistic
clone has a close connection with the problem of identification of a set of
states. That is, Eq. (2.1) is applicable to describe the probabilistic
identification evolution, since $\left\{ \left| \psi _i\right\rangle
^{\otimes \infty },\;i=1,2,...,n\right\} $ are the orthogonal bases of $n$%
-dimension Hilbert space. Inequality (2.3) turns into $X^{(M)}-\Gamma \geq 0$%
, and Inequality (2.4) results in $(\gamma _1+\gamma _2)/2\leq 1-\left|
\left\langle \psi _1|\psi _2\right\rangle \right| ^M$, which is the maximum
identification probability when $n=2$, with $M$ initial copies. In fact,
there is a trade-off between identification and cloning. When the probe
states $\left| P_i\right\rangle $ are orthogonal to each other, we can
identify and clone the input states simultaneously. When $\left|
P_i\right\rangle $ are the same for all the to-be-cloned states, we obtain
no information about the input states and the probabilities of successful
cloning approach the maximum. For a normal situation interpolating between
the cloning and identification, where states $\left| P_i\right\rangle \neq
\left| P_j\right\rangle $ exist, we can identify them with no-zero
probability and get some information about the input, which means the
cloning probabilities must decrease.

Now that the existence of probabilistic cloning and identifying machines has
been demonstrated, the next step is to determine the representations of the
unitary evolution $\hat U$ for the cloning and identifying machines with the
given probability matrix $\Gamma $.

To simplify the deduction, we start with probabilistic identification of one
initial copy. A unitary evolution $\hat U$ is utilized to identify $\left|
\psi _i\right\rangle $, 
\begin{equation}
\hat U\left| \psi _i\right\rangle \left| P_0\right\rangle =\sqrt{\gamma _i}%
\left| \varphi _i\right\rangle \left| P_1\right\rangle +\sum_jc_{ij}\left|
\varphi _j\right\rangle \left| P_0\right\rangle \text{,}  \eqnum{2.5}
\end{equation}
where $\left| P_0\right\rangle $ and $\left| P_1\right\rangle $ are the
orthogonal bases of the probe system. If a postselective measurement of
probe $P$ results in $\left| P_0\right\rangle $, the identification fails.
Otherwise we make a further measurement of the to-be-identified system and
if $\left| \varphi _k\right\rangle $ is detected, the input state should be
identified as $\left| \psi _k\right\rangle $. The inter-inner products of
Eq. (2.5) yield the matrix equation

\begin{equation}
X=\Gamma +CC^{\dagger },  \eqnum{2.6}
\end{equation}

Denoting matrix $A=\left[ \left\langle \varphi _i|\psi _j\right\rangle
\right] _{n\times n}$, we get 
\begin{equation}
X=A^{\dagger }A\text{.}  \eqnum{2.7}
\end{equation}
Obviously $A$ is reversible. Since $\left| \psi _i\right\rangle \left|
P_0\right\rangle =\sum_{m=1}^n\left| \varphi _m\right\rangle \left|
P_0\right\rangle \left\langle \varphi _m|\psi _i\right\rangle $, Eq. (2.5)
can be rewritten as 
\begin{eqnarray*}
&&\hat U\left( \left| \varphi _1\right\rangle \left| P_0\right\rangle
,...,\left| \varphi _n\right\rangle \left| P_0\right\rangle \right) \\
&=&\left( \left| \varphi _1\right\rangle \left| P_1\right\rangle ,...,\left|
\varphi _n\right\rangle \left| P_1\right\rangle \right) \sqrt{\Gamma }A^{-1}
\\
&&+\left( \left| \varphi _1\right\rangle \left| P_0\right\rangle ,...,\left|
\varphi _n\right\rangle \left| P_0\right\rangle \right) C^{\dagger }A^{-1}%
\text{.}
\end{eqnarray*}
On the orthogonal bases $\left\{ \left| \varphi _i\right\rangle \left|
P_j\right\rangle ,i=1,2,...,n,j=0,1\right\} $ in Hilbert space ${\cal H}%
^{AP}={\cal H}\otimes {\cal H}^P$, $\hat U$ can be represented as 
\begin{equation}
U=\left( 
\begin{array}{ll}
C^{\dagger }A^{-1} & M \\ 
\sqrt{\Gamma }A^{-1} & N
\end{array}
\right) \text{,}  \eqnum{2.8}
\end{equation}
where $M,N$ are $n\times n$ matrices. In Appendix A, we derive the
expressions of the four submatrices in Eq. (2.8) and get 
\begin{equation}
U=\tilde VS\tilde V^{\dagger }\text{,}  \eqnum{2.9}
\end{equation}
where $\tilde V=%
\mathop{\rm diag}
\left( V,V\right) $, 
\[
S=\left( 
\begin{array}{ll}
F & -E \\ 
E & F
\end{array}
\right) 
\]
with $E=%
\mathop{\rm diag}
(\sqrt{m_1},...,\sqrt{m_n})$ and $F=%
\mathop{\rm diag}
(\sqrt{1-m_1},...,\sqrt{1-m_n})$. $V$ and $m_i$ are determined by

\begin{equation}
I_n-C^{\dagger }X^{-1}C=Vdiag(m_1,...,m_n)V^{\dagger }.  \eqnum{2.10}
\end{equation}

Since the coefficient matrix $C$ can be deduced from Eq. (2.6), the
parameters $V$ and $m_i,i=1,...,n$ are determined by the probabilities $%
\gamma _i,i=1,...,n$. Hence, the representation $U$ is obtained from the
given probabilities. The expressions of $E$ and $F$ require $0\leq m_i\leq
1, $ $i=1,2,...,n$. In Appendix A we show a more strict limitation $%
0<m_i\leq 1$.

Equation (2.9) is fundamental in obtaining the Hamiltonian and realizing a
quantum probabilistic identifying machine. Based on this representation, we
use the following method to derive the corresponding Hamiltonian. We adopt
the approach in the quantum computation literature of assuming that a
constant Hamiltonian $H$ acts during a short time interval $\Delta t$. Here
we only consider evolution from $t$ to $t+\Delta t$. The time interval is
then related to the strength of couplings in $H$, which are of the order $%
\hbar /\triangle t$. Under this condition we deduce $H$ with 
\begin{equation}
U=e^{-iH\Delta t/\hbar }\text{.}  \eqnum{2.11}
\end{equation}
The unitary representation $U$ in Eq. (2.9) can be diagnolized by
interchanging the columns and rows of the matrix (refer to Appendix B) as

\begin{equation}
U=O%
\mathop{\rm diag}
(e^{i\theta _1},e^{-i\theta _1},...,e^{i\theta _n},e^{-i\theta
_n})O^{\dagger }\text{,}  \eqnum{2.12}
\end{equation}
where $O$ is a unitary matrix and $\theta _j,$ $j=1,...,n$ are determined by

\begin{equation}
e^{i\theta _j}=\sqrt{1-m_j}+i\sqrt{m_j}\text{ }\left( 0<\theta _j\leq \frac 
\pi 2\right) \text{.}  \eqnum{2.13}
\end{equation}
Comparing Eq. (2.12) with Eq. (2.11), the eigenvalues $E_{\pm k}$ of the
Hamiltonians should be 
\begin{equation}
E_{\pm k}=\mp \frac{\theta _k\hbar }{\Delta t}+\frac{2\pi N_{\pm k}\hbar }{%
\Delta t}\text{,}  \eqnum{2.14}
\end{equation}
where $N_{\pm k\text{ }}$are arbitrary integers. $H$ can be represented as 
\begin{equation}
H=O%
\mathop{\rm diag}
(E_1,E_{-1},...,E_n,E_{-n})O^{\dagger }\text{.}  \eqnum{2.15}
\end{equation}

Now we have successfully derived the diagonalized representation and
Hamiltonian of the evolution described by Eq. (2.5), which are essential to
realizing the identification via universal quantum logic gates. We will
extend the result to $M$-initial-copy identification and $M\rightarrow N$
cloning in a similar way. In the situation of probabilistic identification
with $M$ initial copies, we generalize Eq. (2.5) to 
\begin{equation}
\hat U\left| \psi _i\right\rangle ^{\otimes M}\left| P_0\right\rangle =\sqrt{%
\gamma _i}\left| \widetilde{\varphi _i}\right\rangle \left| P_1\right\rangle
+\sum_jC_{ij}\left| \alpha _j\right\rangle \left| P_0\right\rangle \text{,} 
\eqnum{2.16}
\end{equation}
where $\left\{ \left| \widetilde{\varphi _i}\right\rangle
,i=1,2,...,n\right\} $ is a set of orthogonal states in $n^M$-dimensional
Hilbert space ${\cal H}^{\otimes M}$. With the method mentioned above, we
can prove that $U$ has the same representation as that in Eq. (2.9) on
different orthogonal bases $\left\{ \left\{ \left| \alpha _i\right\rangle
\left| P_0\right\rangle \right\} \right. $, $\left. \left\{ \left| 
\widetilde{\varphi _j}\right\rangle \left| P_1\right\rangle \right\}
,i,j=1,2,...,n\right\} $, where the definitions of $V$, $m_i$, $E$, and $F$
are also the same as that of Eq. (2.9). However, they are different in fact
because the determining condition Eq. (2.10) turns into 
\begin{equation}
I_n-C^{\dagger }\left( X^{\left( M\right) }\right) ^{-1}C=V%
\mathop{\rm diag}
(m_1,...,m_n)V^{\dagger }.  \eqnum{2.17}
\end{equation}

As to $M\rightarrow N$ probabilistic cloning, the unitary evolution equation
is Eq. (2.1). Under the same condition of Eq. (2.17) but different
orthogonal bases $\left\{ \left\{ \left| \alpha _i\right\rangle \left|
\varphi _1\right\rangle ^{\otimes (N-M)}\left| P_0\right\rangle \right\} 
\text{, }\left\{ \left| \beta _i\right\rangle \right\} ,i=1,2,...,n\right\} $%
, $U$ may still be represented as that in Eq. (2.9).

We notice that in different situations for probabilistic identification and
cloning the unitary representation and Hamiltonian are of the same form.
However, since the determining conditions are different, the values of $V$, $%
m_i$, $\theta _i,$ and $E_{\pm k}$ are different as well. The unitary
representations and Hamiltonians of different identifications and clones are
based on different bases. All these show that these $\hat U$ or $\hat H$ are
actually different.

In this section, we choose appropriate orthogonal bases and represent the $%
2n^N$-dimensional unitary evolution as Eq. (2.9) in a $2n$-dimensional
subspace. In the subspace orthogonal to such $2n$-dimensional subspace, $U=I$%
.

\section{Networks of probabilistic cloning and identification in a k-partite
system}

So far we have derived the explicit representation of the unitary evolutions
for quantum probabilistic cloning and identification. The next problem is
how to realize these cloning and identifying transformations by physical
means. The fundamental unit of quantum information transmission is the
quantum bit (qubit), i.e., a two-state quantum system, which is capable of
existing in a superposition of Boolean states and of being entangled with
one another. Just as classical bit strings can represent the discrete states
of arbitrary finite dimensionality, a string of $k$ qubits can be used to
represent quantum states in any $2^k$-dimensional Hilbert space. Obviously
there exist $2^k$ linear-independent states in such a $k$-partite system. In
this section we apply the method provided in Sec. II to this special system
and realize probabilistic cloning and identification of an arbitrary state
secretly chosen from a linear-independent state set via universal logic
gates. This solution may be essential to the realization of a quantum
computer.

\subsection{Some basic ideas and notations}

Quantum logic gates have the same number of input and output qubits and a $k$%
-qubit gate carries out a unitary operation of the group $U\left( 2^k\right) 
$, i.e., a generalized rotation in a $2^k$-dimensional Hilbert space. The
formalism we use for quantum computation, which is called a quantum gate
array, was introduced by Deutsch \cite{Deu85}, who showed that a simple
generalization of the Toffoli gate is sufficient as a universal gate for
quantum computation. We introduce this gate as follows.

For any $2\times 2$ unitary matrix 
\[
U=\left( 
\begin{array}{ll}
u_{00} & u_{01} \\ 
u_{10} & u_{11}
\end{array}
\right) 
\]
and $m\in \{0,1,2,...\}$, the matrix corresponding to the $(m+1)$-bit
operation is $\Lambda _m(U)=%
\mathop{\rm diag}
(I_{2^m},U)$, where the bases are lexicographically ordered, i.e., $\left|
000\right\rangle ,\left| 001\right\rangle ,...,\left| 111\right\rangle $.
For a general $U$, $\Lambda _m(U)$ can be regarded as a generalization of
the Toffoli gate, which, on the $m+1$ input bits, applies $U$ to the $(m+1)$%
th bit if and only if the other $m$ bits are all on state $\left|
1\right\rangle $. Barenco {\it et al.} $\ $\cite{Bar95} have further
demonstrated that arbitrary $\Lambda _m(U)$ can be executed by the
combination of a set of one-bit quantum gates $[U(2)]$ and two-bit
Controlled-NOT (C-NOT) gate [that maps Boolean values $(x,y)$ to $(x,x\oplus
y)$].

We first introduce a lemma which shows how to decompose a general unitary
matrix to the product of the matrices $\Lambda _m(U)$.

{\sl Lemma 1.}{\bf \ }Any unitary matrix $U=\left[ u_{ij}\right] _{n\times
n} $ can be decomposed into 
\begin{equation}
U=\left( \prod_{t=1}^{n-1}\prod_{l=t+1}^nA_{tl}\right) \left(
\prod_{k=1}^nB_k\right) ,  \eqnum{3.1}
\end{equation}
where $A_{tl}=[a_{ij\text{ }}^{(tl)}]_{n\times n}$ $=P_{t,n-1}P_{l,n}\hat A(%
\hat u_{tl})P_{l,n}^{\dagger }P_{t,n-1}^{\dagger }$, $B_k=P_{k,n}\hat B%
\left( \exp \left( i\alpha _k\right) \right) P_{k,n}^{\dagger }$, $\hat A(%
\hat u_{tl})=%
\mathop{\rm diag}
(1,1,...,1,\hat u_{tl})$, $\hat u_{tl}$ is a $2\times 2$ unitary matrix, $%
\hat B(\exp (i\alpha _k))=%
\mathop{\rm diag}
(1,1,...,1,\exp (i\alpha _k))$, $P_{ij}$ left-multiplying a matrix
interchanges the $i$th and $j$th row of the matrix, and similarly $%
P_{ij}^{\dagger }$ right-multiplying a matrix interchanges columns. On the
lexicographically ordered orthogonal bases, the representations of operators 
$P_{ij}$ and $P_{ij}^{\dagger }$ are identical. When $n=2^{m+1}$, obviously $%
\hat A(\hat u_{tl})=\Lambda _m(\hat u_{tl})$ and $\hat B\left( \exp \left(
i\alpha _k\right) \right) =\Lambda _m\left( diag\left( 1,\exp \left( i\alpha
_k\right) \right) \right) $.

The meaning of this decomposition in mathematics is that some unitary
matrices, namely $A_{tl}^{\dagger }$, left multiply $U$ to transfer it to a
upper triangular matrix. Since $U$ is unitary, it should be diagonal and can
be decomposed into the product of matrices $B_k$. Thus unitary matrix $U$ is
decomposed into the form of Eq. (3.1).

We show how to transfer the operation $P_{ij}$ to operation $\Lambda
_m(\sigma _x)$ via C-NOT operations. In fact $P_{ij}=\left| \left\{
x_k^i\right\} \right\rangle \left\langle \left\{ x_k^j\right\} \right|
+\left| \left\{ x_k^j\right\} \right\rangle \left\langle \left\{
x_k^i\right\} \right| +\sum_{l\neq i,j}\left| \left\{ x_k^l\right\}
\right\rangle \left\langle \left\{ x_k^l\right\} \right| $, where $\left|
\left\{ x_k^t\right\} \right\rangle =\left|
x_1^t,\;x_2^t,...,\;x_{m+1}^t\right\rangle $ with $x_k^t\in \left\{
0,1\right\} ,\;k=1,2,...,m+1$. These C-NOT operations transfer the subspace
spanned by $\left| \left\{ x_k^i\right\} \right\rangle $ and $\left| \left\{
x_k^j\right\} \right\rangle $ to the subspace spanned by $\left| 11\cdots
11\right\rangle $ and $\left| 11\cdots 10\right\rangle $. For $i\neq j$,
there must exist $k$ satisfying $x_k^i\neq x_k^j$. Denote the minimum value
of $k$ by $k_0$ and assume $x_{k_0}^i=1$, $x_{k_0}^j=0$. For an integer $s$, 
$k_0<s\leq n$, if $x_s^i\neq x_s^j$, we execute C-NOT operation (the $k_0$th
bit controls the $s$th bit). Then for $1\leq h\leq n$, $x_h^i=x_h^j=0$, we
execute $\sigma _x^h$ on the $h$th bit. At last we interchange the input
sequence of the $k_0$th bit and the $(m+1)$th bit.

With the lemma above, a unitary evolution can be expressed as the product of
some controlled unitary operations.

The representations of the input states are another important problem. As to
two linear-independent states $\left| \psi _1\right\rangle ,\left| \psi
_2\right\rangle $ of one qubit system, we set them symmetric, 
\begin{equation}
\left| \psi _{1,2}\left( \theta \right) \right\rangle =\left| \psi _{\pm
}\left( \theta \right) \right\rangle =\cos \theta \left| 0\right\rangle
_A\pm \sin \theta \left| 1\right\rangle _A,  \eqnum{3.2}
\end{equation}
where $0\leq \theta \leq \pi /4$ and $A$ represents the system for
identification and clone. [This simplification is reasonable because
arbitrary states $\left| \psi _1\right\rangle ,\left| \psi _2\right\rangle $
can be transformed to Eq. (3.2) via unitary rotation.]

In the case of a two-partite system, the orthogonal bases are $\left\{
\left| \phi _i\right\rangle _{1,2}\right\} =\left\{ \left| 00\right\rangle
_{1,2},\left| 01\right\rangle _{1,2},\left| 10\right\rangle _{1,2},\left|
11\right\rangle _{1,2}\right\} $ and the input states are $\left\{ \left|
\psi _i\right\rangle _{1,2},\;i=1,2,3,4\right\} $, each of which may be
expressed as $\left| \psi _i\right\rangle _{1,2}=\sum_{j=1}^4t_{ij}\left|
\phi _j\right\rangle _{1,2}$ with $\sum_{j=1}^4\left| t_{ij}\right| ^2=1$.
However, they cannot be converted to symmetric forms as those in Eq. (3.2).
Define $T=[t_{ij}]_{4\times 4}$; the determinant of $T$ should be non-zero
since $\left| \psi _i\right\rangle _{1,2}$ are linear-independent.

{\sl Lemma 2.}{\bf \ }For any four states $\left\{ \left| \psi
_i\right\rangle _{1,2},\;i=1,2,3,4\right\} $ in Hilbert space ${\cal H}%
^{\otimes 2}$ (two-partite system), there exists a unitary operator $U_0$, 
\begin{eqnarray}
&&\ \ U_0\left( \left| \psi _1\right\rangle _{1,2},\;\left| \psi
_2\right\rangle _{1,2},\;\left| \psi _3\right\rangle _{1,2},\;\left| \psi
_4\right\rangle _{1,2}\right)  \eqnum{3.3} \\
\ &=&\left( \left| \phi _1\right\rangle _{1,2},\;\left| \phi _2\right\rangle
_{1,2},\;\left| \phi _3\right\rangle _{1,2},\;\left| \phi _4\right\rangle
_{1,2}\right) \widetilde{T},  \nonumber
\end{eqnarray}
where\widetext
\[
\widetilde{T}=\left( 
\begin{array}{cccc}
1 & e^{i\mu _2^{\left( 1\right) }}\cos \theta _2^{(1)} & e^{i\mu _3^{\left(
1\right) }}\cos \theta _3^{(1)}\cos \theta _3^{(2)} & e^{i\mu _4^{\left(
1\right) }}\cos \theta _4^{(1)}\cos \theta _4^{(2)}\cos \theta _4^{(3)} \\ 
0 & \sin \theta _2^{(1)} & e^{i\mu _3^{\left( 2\right) }}\cos \theta
_3^{(1)}\sin \theta _3^{(2)} & e^{i\mu _4^{\left( 2\right) }}\cos \theta
_4^{(1)}\cos \theta _4^{(2)}\sin \theta _4^{(3)} \\ 
0 & 0 & \sin \theta _3^{(1)} & e^{i\mu _4^{\left( 3\right) }}\cos \theta
_4^{(1)}\sin \theta _4^{(2)} \\ 
0 & 0 & 0 & \sin \theta _4^{(1)}
\end{array}
\right) 
\]
\[
\text{with }0\leq \theta _i^{(1)}<\pi \text{, }0\leq \theta _i^{(j)}<2\pi 
\text{, }0\leq \mu _i^{\left( j\right) }<2\pi \text{.} 
\]
\narrowtext
If $\left\{ \left| \psi _i\right\rangle _{1,2}\right\} $ are
linear-independent, then $\theta _i^{(1)}>0$. The unitarity of $U_0$ yields 
\begin{equation}
T^{\dagger }T=\tilde T^{\dagger }\tilde T.  \eqnum{3.4}
\end{equation}

Lemma 2 can be generalized to $k$-partite system. According to this lemma,
we may concentrate on probabilistic cloning and identification of states $%
\left| \tilde \psi _i\right\rangle _{1,2}=U_0\left| \psi _i\right\rangle
_{1,2}$, $i=1,2,3,4$.

All unitary representations have physical meaning only when the orthogonal
bases have been set. To represent the bases $\left\{ \left| \alpha
_i\right\rangle \right\} $ and $\left\{ \left| \beta _i\right\rangle
\right\} $, we adopt the distinguishability transfer gate ($D$-gate)
operation introduced in Ref. $\cite{Che99}$ and generalize it to a $k$%
-partite system. This operation compresses all the information of the $M$
input copies into one qubit and acts as follows: 
\begin{equation}
D(\theta _1,\theta _2)\left| \psi _{\pm }(\theta _1)\right\rangle \left|
\psi _{\pm }(\theta _2)\right\rangle =\left| \psi _{\pm }(\theta
_3)\right\rangle \left| 0\right\rangle \text{.}  \eqnum{3.5}
\end{equation}

The unitarity of operation $D(\theta _1,\theta _2)$ requires 
\begin{equation}
\cos 2\theta _3=\cos 2\theta _1\cos 2\theta _2.  \eqnum{3.6}
\end{equation}
This condition, together with $0\leq \theta _j\leq \pi /4$, suffices to
determine $\theta _3$ uniquely. Since $D(\theta _1,\theta _2)$ is Hermitian 
\cite{Che99}, it can also transfer state $\left| \psi _{\pm }(\theta
_3)\right\rangle \left| 0\right\rangle $ back to $\left| \psi _{\pm }(\theta
_1)\right\rangle \left| \psi _{\pm }(\theta _2)\right\rangle $. This
accomplishes the process of information decompression. Both the compression
and decompression will be useful in implementing the cloning and
identification.

The $D$-gate operation can be used as an element in a network for $%
M\rightarrow N$ cloning. Define $D_K=D_1\left( \theta _1,\theta
_{K-1}\right) D_2\left( \theta _1,\theta _{K-2}\right) \cdots D_{K-1}(\theta
_1,\theta _1)$, where the operation $D_j\left( \theta _1,\theta
_{K-j}\right) $ compresses the information of partites $j,$ $j+1$ to partite 
$j$, and angles $\theta _{j\text{ }}$ are uniquely determined by $\cos
2\theta _{j+1}=\cos 2\theta _1\cos 2\theta _j$ ($0\leq \theta _j\leq \pi /4$%
). $D_K$ acts as 
\begin{equation}
D_K\left| \psi _{\pm }(\theta _1)\right\rangle ^{\otimes K}=\left| \psi
_{\pm }(\theta _K)\right\rangle _1\left| 0\right\rangle ^{\otimes (K-1)}. 
\eqnum{3.7}
\end{equation}
$\ \ \ \ $ The operations $D_K$, by pairwise interactions, compress all the
information to partite $1$. $D\left( \theta _1,\theta _2\right) $ may be
decomposed into universal operations \cite{Che99}, i.e., local unitary (LU)
operations on a single qubit and C-NOT operations. Here we directly use the
results obtained by Chelfes and Barnett \cite{Che99} and illustrate the D
gate in Fig. 1.

\[
\text{Figure 1.} 
\]

Operation $D_K$ that is suitable for a one-partite system can be generalized
to a $k$-partite system. In the previous part of this subsection we have
discussed the special representations of input states in a $k$-partite
system and we will adopt them below. Consider a two-partite system. Define $%
\tilde \theta =\left\{ \theta _i^{(j)},\;\mu _i^{\left( j\right) },\;2\leq
i\leq 4,\;1\leq j\leq i-1\right\} $ to represent the parameters in matrix $%
\widetilde{T}$ in{\bf \ }lemma 2. We generalize the $D$-gate to two-partite
system, which acts as 
\begin{equation}
D(\tilde \theta ,\tilde \xi )\left| \tilde \psi _i(\tilde \theta
)\right\rangle _A\left| \tilde \psi _i(\tilde \xi )\right\rangle _B=\left| 
\tilde \psi _i(\tilde \eta )\right\rangle _A\left| 00\right\rangle _B, 
\eqnum{3.8}
\end{equation}
where $\tilde \xi $ and $\tilde \eta $ have a definition similar to $\tilde 
\theta $. The unitarity of operation $D(\tilde \theta ,\tilde \xi )$ yields 
\begin{equation}
X(\tilde \theta ,\tilde \xi )=\tilde T^{\dagger }(\tilde \eta )\tilde T(%
\tilde \eta ),  \eqnum{3.9}
\end{equation}
where $X(\tilde \theta ,\tilde \xi )=\left[ _A\left\langle \tilde \psi _i(%
\tilde \theta )\right| \left. \tilde \psi _j(\tilde \theta )\right\rangle
_{AB}\left\langle \tilde \psi _i(\tilde \xi )\right| \left. \tilde \psi _j(%
\tilde \xi )\right\rangle _B\right] _{4\times 4}$. The upper triangular
representation of $\tilde T(\tilde \eta )$ determines $\tilde \eta $
uniquely through Eq. (3.9).

To obtain an explicit expression for the operation $D(\tilde \theta ,\tilde 
\xi )$, we must specify how it transforms states in the subspace orthogonal
to that spanned by $\left| \tilde \psi _i(\tilde \theta )\right\rangle
_A\left| \tilde \psi _i(\tilde \xi )\right\rangle _B$. Equation (3.8) may be
rewritten as 
\begin{eqnarray}
&&\ \ D^{-1}(\tilde \theta ,\tilde \xi )\left\{ \left| \phi _i\right\rangle
_A\left| 00\right\rangle _B,\;i=1,2,3,4\right\}  \eqnum{3.10} \\
\ &=&\left\{ \left| \phi _i\right\rangle _A\left| \phi _j\right\rangle
_B,\;i,j=1,2,3,4\right\} G\tilde T^{-1}(\tilde \eta )\text{,}  \nonumber
\end{eqnarray}
where $G_{16\times 4}$ is the matrix representation of states $\left\{
\left| \tilde \psi _i(\tilde \theta )\right\rangle _A\left| \tilde \psi _i(%
\tilde \xi )\right\rangle _B\right\} $ on the bases $\left\{ \left| \phi
_i\right\rangle _A\left| \phi _j\right\rangle _B\right\} $, which are
lexicographically ordered, i.e., $\left| 0000\right\rangle ,\left|
0001\right\rangle ,...,\left| 1111\right\rangle $ in Hilbert space ${\cal H}%
^{\otimes 2}{\cal \otimes H}^{\otimes 2}$. We denote $G\tilde T^{-1}(\tilde 
\eta )=\left( \omega _1,\omega _5,\omega _9,\omega _{13}\right) $. States $%
\left| \omega _i\right\rangle $ are orthogonal in the space spanned by $%
\left\{ \left| \tilde \psi _i(\tilde \theta )\right\rangle _A\left| \tilde 
\psi _i(\tilde \xi )\right\rangle _B\right\} $. Denote $\tilde G^{-1}=\left(
\omega _1,\omega _2,...,\omega _{16}\right) $, where states $\left\{ \left|
\omega _j\right\rangle ,\;1\leq j\leq 16,\;j\notin \left\{ 1,5,9,13\right\}
\right\} $ are selected in the subspace orthogonal to that spanned by $%
\left\{ \left| \omega _i\right\rangle ,\;i=1,5,9,13\right\} $. With Eq.
(3.10), we let 
\begin{equation}
D^{-1}(\tilde \theta ,\tilde \xi )\left\{ \left| \phi _i\right\rangle
_A\left| \phi _j\right\rangle _B\right\} =\left\{ \left| \phi
_i\right\rangle _A\left| \phi _j\right\rangle _B\right\} \tilde G^{-1}. 
\eqnum{3.11}
\end{equation}
Thus we represent $D(\tilde \theta ,\tilde \xi )$ as matrix $\tilde G$ on
the orthogonal bases $\left\{ \left| \phi _i\right\rangle _A\left| \phi
_j\right\rangle _B\right\} $. Similar to Eq. (3.7), define $D_K=D_1\left( 
\tilde \theta ,\tilde \xi _{K-1}\right) D_2\left( \tilde \theta ,\tilde \xi
_{K-2}\right) \cdots D_{K-1}(\tilde \theta ,\tilde \theta )$ ($\tilde \theta
=\tilde \xi _1$), where $D_j(\tilde \theta ,\tilde \xi )$ compresses the
information of partite systems $A_j,A_{j+1}$ to $A_j$, and $\tilde \xi
_{j+1} $ is uniquely determined by $X(\tilde \theta ,\tilde \xi _j)=\tilde T%
^{\dagger }(\tilde \xi _{j+1})\tilde T(\tilde \xi _{j+1})$. $D_K$ acts as
follows: 
\begin{equation}
D_K\left| \psi _i(\tilde \theta )\right\rangle ^{\otimes K}=\left| \psi _i(%
\tilde \xi _K)\right\rangle _{A_1}\left| 00\right\rangle ^{\otimes (K-1)}. 
\eqnum{3.12}
\end{equation}
We can also define the similar operation $D_K(\tilde \theta ,\tilde \xi )$
that compresses the information of $K$ input copies into one for a $k$%
-partite system, where $\tilde \theta =\left( \theta _i^{\left( j\right)
},\mu _i^{\left( j\right) },i=2,3,...,2^k;j=1,2,...,i-1\right) $. With{\bf \ 
}lemma 1{\bf \ }we can realize $D_K$ via universal logic gates.

For operation $D_K$, we may introduce a new gate called the Controlled-$D_K$
gate, which can transfer the complicated orthogonal bases to
lexicographically ordered ones of a multipartite system. In the information
compression process, we perform a Controlled-$D_K$ gate on the controlled
partites with $P$ as the controller. In the information decompression
process, Controlled-$D_K^{\dagger }$ gate is needed. With all these
operations and controlled operations, we can express the orthogonal bases
and transfer them to those suitable for the realization of quantum cloning
and identification via universal quantum logic gates.

\subsection{Representation of unitary evolution and realization via
universal gates}

Suppose that $\Omega ^k=\left\{ \left| \Omega _i\right\rangle
,i=1,2,...,2^k\right\} $ are the bases which are lexicographically ordered
in Hilbert space ${\cal H}^{\otimes k}$. For the given probability matrix $%
\Gamma $, with a $D_K$ gate, we can represent the orthogonal bases $%
\{\left\{ \left| \alpha _i\right\rangle \left| P_0\right\rangle \right\} $, $%
\left\{ \left| \widetilde{\varphi _j}\right\rangle \left| P_1\right\rangle
\right\} $, $i$, $j=1,2\ldots ,2^k\}$ [of Eq. (2.16) for probabilistic
identification] and $\left\{ \left\{ \left| \alpha _i\right\rangle \left|
\varphi _1\right\rangle ^{\otimes (N-M)}\left| P_0\right\rangle \right\} 
\text{, }\left\{ \left| \beta _j\right\rangle \right\}
,i,j=1,2,...,2^k\right\} $ [of Eq. (2.1) for probabilistic cloning] as 
\begin{equation}
\begin{array}{c}
\left\{ \left\{ D_M^{-1}\left| \Omega _i\right\rangle \left| \Omega
_1\right\rangle ^{\otimes (M-1)}\left| P_0\right\rangle \right\} ,\right. \\ 
\text{ }\left. \left\{ \left| \Omega _j\right\rangle \left| \Omega
_1\right\rangle ^{\otimes (M-1)}\left| P_1\right\rangle \right\} ,\text{ }%
i,j=1,2,...,2^k\right\} \\ 
\\ 
\left\{ \left\{ D_M^{-1}\left| \Omega _i\right\rangle \left| \Omega
_1\right\rangle ^{\otimes (N-1)}\left| P_0\right\rangle \right\} ,\right. 
\text{ } \\ 
\left. \text{ }\left\{ D_N^{-1}\left| \Omega _j\right\rangle \left| \Omega
_1\right\rangle ^{\otimes (N-1)}\left| P_1\right\rangle \right\} ,\text{ }%
i,j=1,2,...,2^k\right\}
\end{array}
,  \eqnum{3.13}
\end{equation}
where the first expression is for identification and the second is for
cloning. With a controlled-$D_M$ gate and a controlled-$D_N$ gate, we can
transfer these orthogonal bases into 
\begin{eqnarray}
&&\ \ \ \left\{ \left\{ \left| \Omega _i\right\rangle _{A_1}\left|
P_0\right\rangle \right\} ,\left\{ \left| \Omega _j\right\rangle
_{A_1}\left| P_1\right\rangle \right\} ,\text{ }i,j=1,2,...,2^k\right\} 
\eqnum{3.14} \\
&&\ \ \ \otimes \left| \Omega _1\right\rangle _{A_2,A_3,...,A_K}^{\otimes
(K-1)}  \nonumber
\end{eqnarray}
where $K=M$ is for identification and $K=N$ is for cloning. On these new
orthogonal bases, the evolution $\hat U$ is a unitary controlled operation
on a composite system of $A_1$ and probe $P$ with the composite system of
subsystem $A_2,A_3,...,A_K$ as the controller. If the controller is in state 
$\left| \Omega _1\right\rangle _{A_2,A_3,...,A_K}^{\otimes (K-1)}$, we
perform operation $\hat U$ on the composite system of $A_1P$. Otherwise we
make no operation. Denote $\left| P_0\right\rangle =\left| 0\right\rangle _P$%
, $\left| P_1\right\rangle =\left| 1\right\rangle _P$, on the bases $\left\{
\left\{ \left| \Omega _i\right\rangle _{A_1}\left| 0\right\rangle _P\right\}
,\left\{ \left| \Omega _j\right\rangle _{A_1}\left| 1\right\rangle
_P\right\} \right. $, $\left. i,j=1,2,\ldots ,2^k\right\} $; $U$ can be
represented as $U=\tilde VS\tilde V^{\dagger }$ [Eq. (2.9)]. $\widetilde{V}$
corresponds to operation 
\begin{equation}
\hat V_{A_1}\hat I_P\left| \Omega _1\right\rangle ^{\otimes (K-1)\ \otimes
(K-1)}\left\langle \Omega _1\right| +\hat J,  \eqnum{3.15}
\end{equation}
where $\hat J=\sum_{\left| \left\{ \Omega _{i_j}\right\} \right\rangle \neq
\left| \Omega _1\right\rangle ^{\otimes (K-1)}}\hat I_{A_1P}\left| \left\{
\Omega _{i_j}\right\} \right\rangle \left\langle \left\{ \Omega
_{i_j}\right\} \right| $ with $\left| \left\{ \Omega _{i_j}\right\}
\right\rangle =\left| \Omega _{i_1}\right\rangle \left| \Omega
_{i_2}\right\rangle \cdots \left| \Omega _{i_{K-1}}\right\rangle $, $K=M$ is
for identification, and $K=N$ is for cloning. The matrix corresponding to
the operation $\hat V_{A_1}$ on the bases $\left\{ \left| \Omega
_i\right\rangle _{A_1}\right\} $ is $V$. $\hat I_P$ represents unit
operation of a probe system. On the new orthogonal bases $\left\{ \left\{
\left| \Omega _i\right\rangle _{A_1}\left| P_0\right\rangle ,\;\left| \Omega
_i\right\rangle _{A_1}\left| P_1\right\rangle \right\}
,\;i=1,2,...,2^k\right\} $, we express 
\[
S=\left( 
\begin{array}{cc}
F & -E \\ 
E & F
\end{array}
\right) 
\]
as $S=%
\mathop{\rm diag}
(K_1,K_2,...,K_{2^k})$, where 
\[
K_i=\left( 
\begin{array}{cc}
\sqrt{1-m_i} & -\sqrt{m_i} \\ 
\sqrt{m_i} & \sqrt{1-m_i}
\end{array}
\right) . 
\]
So we obtain 
\begin{eqnarray}
\hat S &=&\prod_{i=1}^{2^k}P_{2i,2^{k+1}}P_{2i-1,2^{k+1}-1}\Lambda
_k^{A_1P}(K_i)  \eqnum{3.16} \\
&&\ \times P_{2i-1,2^{k+1}-1}P_{2i,2^{k+1}}\left| \Omega _1\right\rangle
^{\otimes (K-1)\ \otimes (K-1)}\left\langle \Omega _1\right| +\hat J, 
\nonumber
\end{eqnarray}
where $K=M$ is for identification and $K=N$ is for cloning. We have shown in%
{\bf \ }lemma 1{\bf \ }that the unitary operations $U_0$, $%
P_{2i-1,2^{k+1}-1} $, $P_{2i,2^{k+1}},$ and $\hat V_{A_1}$ can be decomposed
into the product of basis operations such as C-NOT and $\Lambda _k(\hat u)$.
The decomposition of $\Lambda _k(\hat u)$ has been completed by Barenco {\it %
et al.} $\ $\cite{Bar95}. Thus we complete the decomposition of the unitary
evolution via universal quantum logic gates, so as to realize probabilistic
cloning and identification of an $n$-partite system.

In the following we will give some examples. First we shall be concerned
with quantum probabilistic identification of a one-partite system, provided
with $M$ initial copies. With the given maximum probability $\gamma
_1=\gamma _2=1-\cos ^M2\theta $, we obtain 
\[
\widetilde{V}=\frac 1{\sqrt{2}}(I^A+i\sigma _y^A)I^P=R_y^A(\frac \pi 2%
)I^P,m_1=1, 
\]
\[
m_2=\frac{1-\cos ^M2\theta }{1+\cos ^M2\theta },K_1=\left( 
\begin{array}{cc}
0 & -1 \\ 
1 & 0
\end{array}
\right) , 
\]
\[
K_2=\left( 
\begin{array}{ccc}
\sqrt{\frac{2\cos ^M2\theta }{1+\cos ^M2\theta }} &  & -\sqrt{\frac{1-\cos
^M2\theta }{1+\cos ^M2\theta }} \\ 
&  &  \\ 
\sqrt{\frac{1-\cos ^M2\theta }{1+\cos ^M2\theta }} &  & \sqrt{\frac{2\cos
^M2\theta }{1+\cos ^M2\theta }}
\end{array}
\right) , 
\]
where 
\[
R_y(\chi )=\left( 
\begin{array}{ll}
\cos \frac \chi 2 & \sin \frac \chi 2 \\ 
-\sin \frac \chi 2 & \cos \frac \chi 2
\end{array}
\right) . 
\]
The network of quantum probabilistic identification for a one-partite system
via universal logic gates is shown in Fig. 2 ($M=2$). 
\[
\text{Figure 2.} 
\]

The S gate in Fig. 2 is illustrated in Fig. 3. 
\[
\text{Figure 3.} 
\]

For a two-partite system, with the given maximum probability matrix $\Gamma $
which satisfies the inequality $X^{\left( M\right) }-\Gamma \geq 0$, we
obtain 
\begin{eqnarray*}
\hat S &=&\sigma _x^1\sigma _x^2\Lambda _2(K_1)\sigma _x^2\sigma _x^1\sigma
_x^1\Lambda _2(K_2)\sigma _x^1 \\
&&\times \sigma _x^2\Lambda _2(K_3)\sigma _x^2\Lambda _2(K_4)\left|
00\right\rangle ^{\otimes (M-1)\ \otimes (M-1)}\left\langle 00\right| +\hat J%
.
\end{eqnarray*}
The network of quantum probabilistic identification for a two-partite system
is shown in Fig. 4 ($M=2$). 
\[
\text{Fig. 4.} 
\]
The S-gate in Figure 4 is illustrated in Fig. 5 
\[
\text{Fig. 5.} 
\]

As to probabilistic cloning, we also begin with a one-partite system. With
inequality (2.4), we give the maximum probability $\gamma _{\max }=\left(
1-\cos ^M2\theta \right) /\left( 1-\cos ^N2\theta \right) $. Then 
\begin{eqnarray*}
\widetilde{V} &=&\frac 1{\sqrt{2}}(I^{A_1}+i\sigma _y^{A_1})\left|
0\right\rangle _{A_2A_2}\left\langle 0\right| I^P \\
\ &=&R_y^A(\frac \pi 2)\left| 0\right\rangle _{A_2A_2}\left\langle 0\right|
I^P,
\end{eqnarray*}
\[
K_1=\left( 
\begin{array}{ll}
0 & -1 \\ 
1 & 0
\end{array}
\right) , 
\]
\widetext
\[
K_2=\left( 
\begin{array}{ccc}
\sqrt{\frac{2\left( \cos ^M2\theta -\cos ^N2\theta \right) }{\left( 1-\cos
^N2\theta \right) \left( 1+\cos ^M2\theta \right) }} &  & -\sqrt{\frac{%
\left( 1+\cos ^N2\theta \right) \left( 1-\cos ^M2\theta \right) }{\left(
1-\cos ^N2\theta \right) \left( 1+\cos ^M2\theta \right) }} \\ 
&  &  \\ 
\sqrt{\frac{\left( 1+\cos ^N2\theta \right) \left( 1-\cos ^M2\theta \right) 
}{\left( 1-\cos ^N2\theta \right) \left( 1+\cos ^M2\theta \right) }} &  & 
\sqrt{\frac{2\left( \cos ^M2\theta -\cos ^N2\theta \right) }{\left( 1-\cos
^N2\theta \right) \left( 1+\cos ^M2\theta \right) }}
\end{array}
\right) . 
\]
\narrowtext
The network of quantum probabilistic cloning for a one-partite system is
shown in Fig. 6 ($M=2$, $N=3$).

\[
\text{Fig. 6. } 
\]

For a two-partite system, with the given maximum probability matrix $\Gamma $
satisfying $X^{\left( M\right) }-\sqrt{\Gamma }X^{\left( N\right) }\sqrt{%
\Gamma }\geq 0$, we obtain 
\begin{eqnarray*}
\hat S &=&\sigma _x^1\sigma _x^2\Lambda _2(K_1)\sigma _x^2\sigma _x^1\sigma
_x^1\Lambda _2(K_2)\sigma _x^1 \\
&&\times \sigma _x^2\Lambda _2(K_3)\sigma _x^2\Lambda _2(K_4)\left|
00\right\rangle ^{\otimes (N-1)\ \otimes (N-1)}\left\langle 00\right| +\hat J%
.
\end{eqnarray*}
The network of quantum probabilistic cloning for a two-partite system is
shown in Fig. 7 (where $M=2$, $N=3$).\quad 
\[
\text{Fig. 7. } 
\]
\quad

So far we have realized quantum probabilistic identification and cloning in
a $k$-partite system via universal quantum logic gates, which have important
applications in quantum cryptography \cite{Ben92,Bar93}, quantum programming 
\cite{Nie97}, and quantum state preparation \cite{Bru92}.

\subsection{Robustness of the quantum networks}

The robustness properties of the cloning and identifying machines may prove
to be crucial in practice. In this subsection, we show whether any errors
occur in the input target systems $A_{M+1},$ $A_{M+2},...,A_N$, we can
detect them without destroying the to-be-cloned states in systems $A_1,$ $%
A_2,...,A_M$, and the to-be-cloned states can be recycled.

The input target state with errors may be generally expressed as 
\begin{eqnarray}
&&\ \rho _{A_{M+1},A_{M+2},...,A_N}  \eqnum{3.17} \\
\ &=&\left( \left( 1-\delta _1\right) \left| \Omega _1\right\rangle
\left\langle \Omega _1\right| +\delta _1\sum\limits_{i=2}^{2^k}\epsilon
_i\left| \Omega _i\right\rangle \left\langle \Omega _i\right| \right)
^{\otimes (N-M)}\text{,}  \nonumber
\end{eqnarray}
where $\sum_{i=2}^{2^k}\left| \epsilon _i\right| =1$ and $\delta _1$ is the
error rate, or 
\begin{eqnarray}
&&\ \left| \phi \right\rangle _{A_{M+1},A_{M+2},...,A_N}  \eqnum{3.18} \\
\ &=&\left( \sqrt{1-\left| \delta _2\right| ^2}\left| \Omega _1\right\rangle
+\delta _2\sum\limits_{i=2}^{2^k}\tau _i\left| \Omega _i\right\rangle
\right) ^{\otimes (N-M)}  \nonumber
\end{eqnarray}
where $\sum_{i=2}^{2^k}\left| \tau _i\right| ^2=1$ and $\left| \delta
_2\right| ^2$ is the error rate. Equation (3.17) expresses the errors caused
by the decoherence due to the environment. Equation (3.18) represents the
errors in state preparation. The errors occur in the $\left( N-M\right) $
input target systems for cloning with the approximate rate $\left(
N-M\right) \delta _1$ [$\left( N-M\right) \left| \delta _2\right| ^2$],
which cannot be omitted in practice when $N$ is relatively large.

After the cloning process, if measurement of probe $P$ results in $\left|
0\right\rangle _P$, the cloning attempt should be regarded as a failure in a
normal sense. However, it may be caused by errors.

If errors caused by the decoherence occur in any input target systems, at
least one system occupies state $\left| \Omega _i\right\rangle $, $i\neq 1$.
According to Eqs. (3.15) and (3.16), the controlled operations $\hat V_{A_1}$%
, $\hat S$, and Controlled-$D_K^{\dagger }$ gate in the information
decompression, function as unit evolutions, in other words, only Controlled-$%
D_K$ gate in the information compression works. Thus the to-be-cloned state
remains undestroyed. According to Eq. (2.1) and the above discussion, the
input target states remain unchanged if probe $P$ is in $\left|
0\right\rangle _P$, whenever the clone fails or errors occur. These two
cases can be checked out by measuring the output target states.

If the errors are caused by state preparation, after the evolution of the
system, the output target system corresponding to $\left| 0\right\rangle _P$
is the superposition of two different terms. We measure the output target
states, and if they result in $\left| \Omega _1\right\rangle ^{\otimes
(N-M)} $, the clone really fails. Otherwise, the errors work and the
to-be-cloned state remains undestroyed.

To the two error situations mentioned above, we can reinput the to-be-cloned
system to the cloning machines at the location immediately behind the
Controlled-$D_K$ gate (the first operation of the cloning machine) and clone
again.

\section{Conclusions}

In summary, we have considered the realization of quantum probabilistic
identifying and cloning machines by physical means. We showed that the
unitary representation and the Hamiltonian of probabilistic cloning and
identifying machines are determined by the probabilities of success. The
logic networks have been obtained by decomposing the unitary representation
into universal quantum logic operations. We have discussed the robustness of
the networks and found that if error occurs in the input target system, we
can detect it and the to-be-cloned states can be recycled. Our method is
suitable for $k$-partite system, such as a quantum computer, and may be
generalized to general state-dependent cloning and identification.

\begin{center}
{\bf ACKNOWLEDGMENTS}
\end{center}

We thank Dr. L.-M. Duan for helpful discussion. This work was supported by
the National Natural Science Foundation of China.

\begin{center}
\appendix{\bf APPENDIX A}
\end{center}

In this appendix, we determine $M$ and $N$ and derive the representation of $%
U$. $U$ is a unitary matrix, that is, 
\begin{equation}
UU^{\dagger }=U^{\dagger }U=I_{2n}\text{.}  \eqnum{A1}
\end{equation}

Eq. (A1) can be proved equivalent to two equations below 
\begin{equation}
N=-(\sqrt{\Gamma })^{-1}CM\text{,}  \eqnum{A2}
\end{equation}
\begin{equation}
MM^{\dagger }=I_n-C^{\dagger }X^{-1}C\text{.}  \eqnum{A3}
\end{equation}

It is obvious that $I_n-C^{\dagger }X^{-1}C$ is a symmetric matrix.
According to Eq. (2.6), we yield 
\begin{equation}
I_n-C^{\dagger }X^{-1}C=(I_n+C^{\dagger }\Gamma ^{-1}C)^{-1}\text{.} 
\eqnum{A4}
\end{equation}

For $\Gamma $ positive definite, $C^{\dagger }\Gamma ^{-1}C$ is semipositive
definite. Thus $I_n+C^{\dagger }\Gamma ^{-1}C$ is positive definite and its
reversed matrix $I_n-C^{\dagger }X^{-1}C$ is also positive definite.

$I_n-C^{\dagger }X^{-1}C$ can be represented as the following: 
\begin{equation}
I_n-C^{\dagger }X^{-1}C=Vdiag(m_1,...,m_n)V^{\dagger }\text{,}  \eqnum{A5}
\end{equation}
where $V$ is unitary. Together with Eq. (A3), $M$ is determined by 
\begin{equation}
M=-V%
\mathop{\rm diag}
(\sqrt{m_1},...,\sqrt{m_n})V^{\dagger }\text{.}  \eqnum{A6}
\end{equation}

Furthermore, we can also prove several useful conclusions to replace the
submatrices of $U$ in Eq. (2.8), 
\begin{equation}
C^{\dagger }A^{-1}=V%
\mathop{\rm diag}
(\sqrt{1-m_1},...,\sqrt{1-m_n})V^{\dagger }\text{,}  \eqnum{A7}
\end{equation}
\begin{equation}
\sqrt{\Gamma }A^{-1}=V%
\mathop{\rm diag}
(\sqrt{m_1},...,\sqrt{m_n})V^{\dagger },  \eqnum{A8}
\end{equation}
\begin{equation}
N=-(\sqrt{\Gamma })^{-1}CM=V(\sqrt{1-m_1},...,\sqrt{1-m_n})V^{\dagger }\text{%
.}  \eqnum{A9}
\end{equation}

Hence, we get 
\begin{equation}
U=\left( 
\begin{array}{ll}
V & 0 \\ 
0 & V
\end{array}
\right) \left( 
\begin{array}{ll}
F & -E \\ 
E & F
\end{array}
\right) \left( 
\begin{array}{ll}
V^{\dagger } & 0 \\ 
0 & V^{\dagger }
\end{array}
\right) \text{,}  \eqnum{A10}
\end{equation}
where $E=%
\mathop{\rm diag}
(\sqrt{m_1},...,\sqrt{m_n}),$ $F=%
\mathop{\rm diag}
(\sqrt{1-m_1},...,\sqrt{1-m_n})$.

According to Eq. (A5) and $I_n-C^{\dagger }X^{-1}C>0$, we yield 
\[
m_i>0,\qquad \text{ }i=1,2,...,n\text{.} 
\]

On the other hand, Eq. (A5) can be rewritten as 
\[
C^{\dagger }X^{-1}C=V%
\mathop{\rm diag}
(1-m_1,...,1-m_n)V^{\dagger }\text{.} 
\]

For $X$ positive definite, $C^{\dagger }X^{-1}C$ is semipositive definite.
So $1-m_i\geq 0$, that is, $m_i\leq 1,$ $i=1,...,n$. Combining the results
above, we get the range of $m_i$ as 
\begin{equation}
0<m_i\leq 1,\text{ }i=1,2,...,n\text{.}  \eqnum{A11}
\end{equation}

\begin{center}
\appendix{\bf APPENDIX B}
\end{center}

Here we diagonalize $\hat U$.

$S$ can be rewritten as 
\begin{equation}
S=TKT^{\dagger }\text{,}  \eqnum{B1}
\end{equation}
where $K=diag(K_1,K_2,...,K_n)$, 
\[
K_i=\left( 
\begin{array}{ll}
\sqrt{1-m_i} & -\sqrt{m_i} \\ 
\sqrt{m_i} & \sqrt{1-m_i}
\end{array}
\right) , 
\]
$T$ is a unitary matrix which interchanges the rows of $K$, and $T^{\dagger
} $ interchanges the columns. Denoting 
\[
L_i=\frac 1{\sqrt{2}}\left( 
\begin{array}{ll}
1 & -i \\ 
-i & 1
\end{array}
\right) , 
\]
$i=1,2,...,n,$ $\tilde L=%
\mathop{\rm diag}
(L_1,L_2,...,L_n)$, we have 
\begin{equation}
K=\tilde L%
\mathop{\rm diag}
(e^{i\theta _1},e^{-i\theta _1},...,e^{i\theta _n},e^{-i\theta _n})\tilde L%
^{\dagger }\text{,}  \eqnum{B2}
\end{equation}
where $\theta _j$ ( $j=1,2,...,n)$ are determined by 
\begin{equation}
e^{i\theta _j}=\sqrt{1-m_j}+i\sqrt{m_j},\qquad \text{ }\left( 0<\theta
_j\leq \frac \pi 2\right) \text{.}  \eqnum{B3}
\end{equation}

According to Eqs. (2.9), (B1), and (B2), $U$ is completely diagonalized as
the following: 
\begin{equation}
U=O%
\mathop{\rm diag}
(e^{i\theta _1},e^{-i\theta _1},...,e^{i\theta _n},e^{-i\theta
_n})O^{\dagger }\text{,}  \eqnum{B4}
\end{equation}
where $O=\tilde VT\tilde L$.

\begin{center}
{\bf FIGURE\ CAPTIONS}
\end{center}

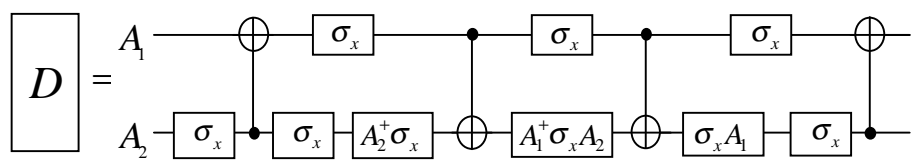
FIG. 1. The networks of a $D$ gate \cite{Che99}. $\bullet $ and $\oplus $
denote the controller and target bit of a C-NOT operation, respectively.

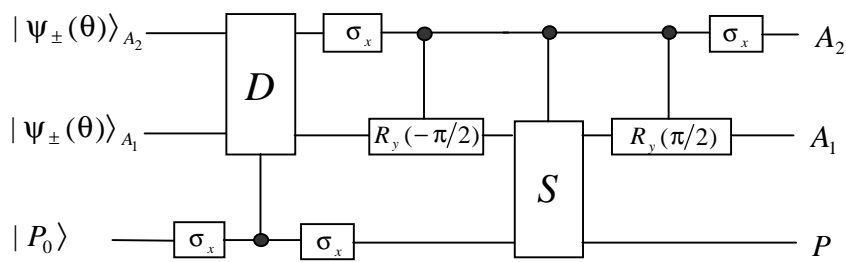
FIG. 2. The networks of probabilistic identification for a one-partite
system. $\left| \psi _{\pm }\left( \theta \right) \right\rangle _{A_i}$ are
to-be-identified states and $\left| P_0\right\rangle $ is the probe.

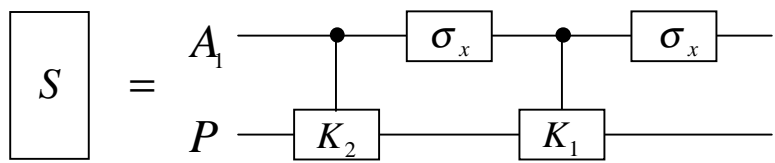
FIG. 3. The networks of an $S$ gate for a one-partite system.

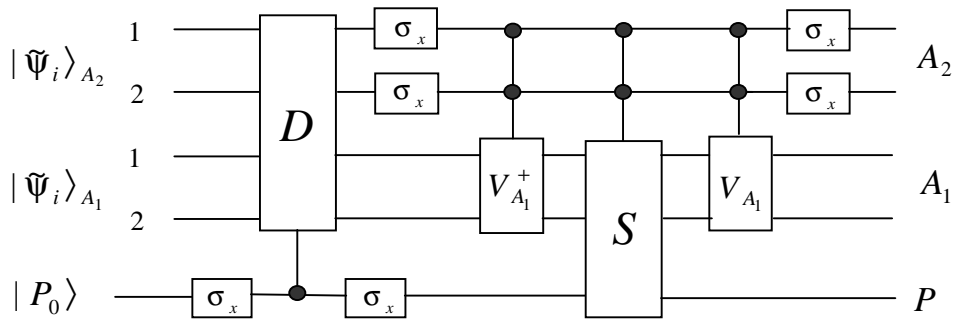
FIG. 4. The networks of probabilistic identification for a two-partite
system. $\left| \tilde \psi _i\right\rangle _{A_j}$ are to-be-identified
states.

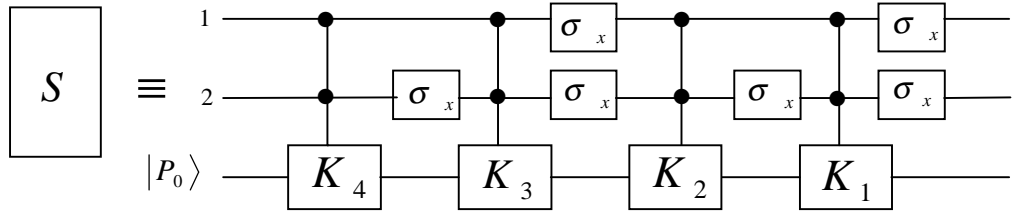
FIG. 5. The networks of an $S$ gate for a two-partite system.

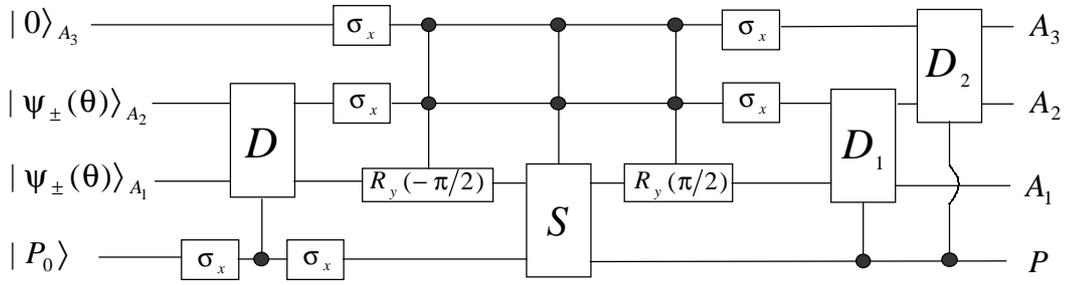
FIG. 6. The networks of probabilistic cloning of a one-partite system. The $%
S $ gate has been illustrated in Fig. 3. $\left| \psi _{\pm }\left( \theta
\right) \right\rangle _{A_i}$ are to-be-cloned states and $\left|
0\right\rangle _{A_3}$ is the input target state.

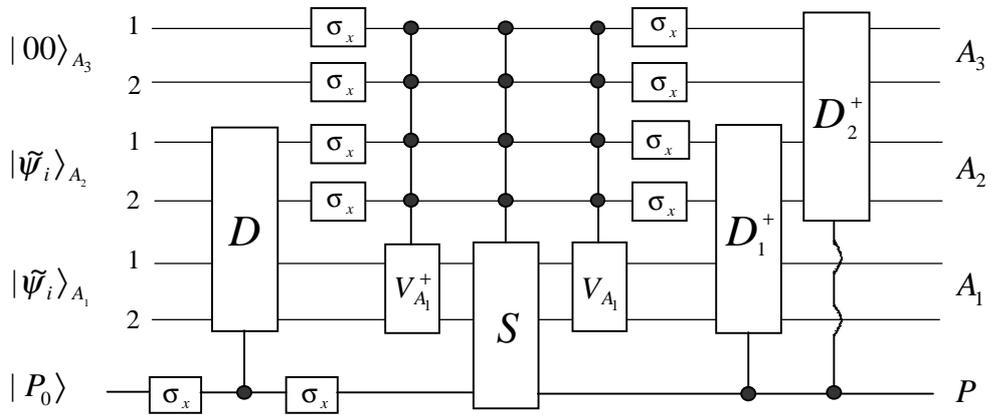
FIG. 7. The networks of probabilistic cloning for a two-partite system. The $%
S$ gate has been illustrated in Fig. 5. $\left| \tilde \psi _i\right\rangle
_{A_j}$ are to-be-cloned states and $\left| 00\right\rangle _{A_3}$ is the
input target state.

\widetext


\begin{references}
\bibitem{Woot82}  W. K. Wootters and W. H. Zurek, Nature (London) {\bf 299},
802 (1982).

\bibitem{Yuen86}  H. P. Yuen, Phys. Lett. A {\bf 113}, 405 (1986).

\bibitem{Yuen96}  G. M. D'Ariano and H. P. Yuen, Phys. Rev. Lett. {\bf 76},%
{\bf \ }2832 (1996).

\bibitem{Bar96}  H. Barnum, C. M. Caves, C. A. Fuchs, R. Jozsa, and B.
Schumacher, Phys. Rev. Lett. {\bf 76},{\bf \ }2818 (1996).

\bibitem{Koa98}  M. Koashi and N. Imoto, Phys. Rev. Lett. {\bf 81},{\bf \ }%
4264 (1998).

\bibitem{Die82}  D. Dieks, Phys. Lett. {\bf 92A}, 271 (1982); Phys. Lett. A 
{\bf 126},{\bf \ }303 (1987).

\bibitem{Ben95}  C. H. Bennett, Phys. Today {\bf 48},{\bf \ }24 (1995).

\bibitem{Buz96}  V. Bu\u zek and M. Hillery,\ Phys. Rev. A {\bf 54}, 1844
(1996).

\bibitem{Gis97}  N. Gisin and S. Massar, Phys. Rev. Lett. {\bf 79}, 2153
(1997).

\bibitem{Bru98A}  D.Bru$\beta $, D. P. DiVincenzo, A. Ekert, C. A. Fuchs, C.
Macchiavello, and J. A. Smolin, Phys. Rev. A {\bf 57}, 2368 (1998).

\bibitem{Bru98L}  D.Bruss, A. Ekert, and C. Macchiavello,\ Phys. Rev. Lett. 
{\bf 81}, 2598 (1998).

\bibitem{Wer98}  R. F. Werner,\ Phys. Rev. A {\bf 58}, 1827 (1998).

\bibitem{Key99}  M. Keyl and R. F. Werner, J. Math. Phys. {\bf 40}, 3283
(1999).

\bibitem{Mas95}  S. Massar and S. Popescu, Phys. Rev. Lett. {\bf 74},{\bf \ }%
1259 (1995).

\bibitem{Der98}  R. Derka, V. Bu\v zek, and A. Ekert, Phys. Rev. Lett. {\bf %
80},{\bf \ }1571 (1998).

\bibitem{Hil98}  M. Hillery and V. Bu\u zek, Phys. Rev. A {\bf 56}, 1212
(1997).

\bibitem{Che99}  A. Chefles and S. M. Barnett, Phys. Rev. A {\bf 60},{\bf \ }%
136 (1999).

\bibitem{Dua98A}  L.-M. Duan and G.-C. Guo, Phys. Lett. A {\bf 243},{\bf \ }%
261 (1998).

\bibitem{Dua98L}  L.-M. Duan and G.-C. Guo, Phys. Rev. Lett. {\bf 80}, 4999
(1998).

\bibitem{Zha99}  C.-W. Zhang, C.-F. Li, and G.-C. Guo, Phys. Lett. A {\bf 261%
}, 25 (1999).\ 

\bibitem{Buz97}  V. Bu\v zek, S. L. Braunstein, M. Hillery, and D. Bru$\beta 
$, Phys. Rev. A {\bf 56}, 3446 (1997).

\bibitem{Deu85}  D. Deutsch, Proc. R. Soc. London, Ser. A {\bf 400}, 97
(1985), {\bf 425}, 73 (1989).

\bibitem{Bar95}  A. Barenco, C. H. Bennett, R. Cleve, D. P. DiVincenzo, N.
Margolus, P. Shor, T. Sleator, J. A. Smolin, and H. Weinfurter,\ Phys. Rev.
A {\bf 52}, 3457 (1995).

\bibitem{Ben92}  C. H. Bennett, G. Brassard, and N. D. Mermin,\ Phys. Rev.
Lett. {\bf 68}, 557 (1992).

\bibitem{Bar93}  S. M. Barnett and S. J. D. Phoenix, Phys. Rev. A {\bf 48},
R5 (1993).

\bibitem{Nie97}  M. A. Nielsen and I. L. Chuang,\ Phys. Rev. Lett. {\bf 79},
321 (1997).

\bibitem{Bru92}  M. Brune, S. Haroche, J. M. Raimond, L. Davidovich, and N.
Zagury, Phys. Rev. A {\bf 45}, 5193 (1992).
\end{references}
\end{document}